# The informational physics indeed can help to understand Nature?

S.V. Shevchenko, V.V. Tokarevsky

*Abstract*

In our previous articles ("The Information and the Matter", v1, v5; more systematically the informational conception is presented in arXiv paper "The Information as Absolute", 2010) it was rigorously shown that the Matter is some informational system (structure), which is an [practically] infinitesimal sub-sets of the utmost infinite and fundamental Set "Information". Here some physical consequences from this conception are considered.

## 1. INTRODUCTION

In the Refs. [1] - [4] it was rigorously logically shown that the entity/ concept "Information" is utmost general and fundamental when all/anything what exist is/are some realizations of the information – all/anything what exist is/are "the words", some elements of utmost general and fundamental infinite Set "Information". Suggested concept makes more clear a number of ontological and epistemological problems in science, first of all – the *problem of cognition,* i.e. the problem of adequacy of a human's consciousness inferences (in form of some language statements, including mathematical and algorithmic languages) to the reality, - becomes much more understandable since the elements of the Set "Information" are some information statements also.

This Set /concept/entity has very unusual and interesting properties, including that the entity and the Set are the same, so both above are entitled identically. Besides note some another important properties of the Set (more about the concept see [4]).

Just as a consequence of ultimate fundamentality of the Set "Information" it turns out to be possible to prove *completely* the *existence* and the *truth* of the concept, unlike to any other concept, e.g. – to the physical concepts and laws. Indeed - note that – what follows also from Gedel's theorems – *any indeed new* concept/law can be obtained by human consciousness *only at some experiment*; any actually new information for the Consciousness is totally empirical. This new knowledge human's consciousness generalises and states as some system of axioms; in Nature sciences – as the system of "Nature laws". Further – *any logical sequences from defined system of axioms*, i.e. – theorems, calculations, etc. ***don't create any new information*** in addition to the information that *exists already* in some *implicit forms* in the axioms.

The Information concept is rigorously logically grounded and is completely self-consistent. Note, however, that just because of it's utmost fundamentality and self-consistency we can not go out from the concept to prove concept's *uniqueness*.

The main epistemological problem now is that any new concept/law in Nature sciences, besides the Information concept, e.g., any physical law, *can not be proven at all* – all the laws are empirical inferences *grounded* by using only *necessary but non – sufficient criterion* on the reiteration of experimental results. The Information concept also should be detected only at an experiment, as some information about existence of some language and a number of logical rules, but in this case *it is sufficient* "to make only one experiment" to obtain/ to prove further the existence of the Information concept as something utmost fundamental as well as a number of it's [Set's] properties.

Another property of the Information concept is that any information is some "generalization" and always *implicitly* contains, besides some specific data in specific context, "compressed" information about these data in any possible contexts. Some example of an implicit information is given above – all/ any data, which can be obtained for a system of axioms as any logical/ mathematical consequences from these axioms, don't contain any new information, additional to the information that the axioms contain already.

Moreover – *any particular "notion"* (e.g. – material thing), on another words – any element in the Set, *contains all infinite true information which exists in the Set* as information negation which logically differs (singles out) this element from the rest in the Set and so any element in the Set exists always as a bit "I/not I", so all Set formally can be reduced to *any single* It's element completely. Including – to "Zero element" ("null set") of the Set or to the information which, in principle, can exist when there exist no anything else – and this information can be expressed in maybe any existent language as infinite cyclic statement "there is no anything besides the information that there is no anything, besides…".

The text above contains, however, some logical uncertainty: the Information is defined as a concept and as a set, when there exists no definition of the set till now. The definition problem for the set is similar to the information definition problem considered in [1] – any invented definition turns out to be a tautology: "the set is the set" ("collection", "ensemble", "manifold", etc.). And again, as that was in the case of the information definition problem, one can say that this "many years experiment", which resulted in the absence of the set's definition, shows that the set concept is – like to the information concept - something utmost fundamental.

From above it follows, that the Set "Information" (and every element in the Set) "contained", contains and "will contain" all information about all in "all times", i.e. It is simultaneously "static" and "dynamic» Set. "Static" – since it doesn't change because is always absolutely totally complete, "dynamic" – since it contains some dynamic elements; e.g. – any possible logical sequences (including, for example, creation and evolution of our Universe).

So it would not be too surprised that the Set "Information" is something like as Cantor's "Absolute Infinite" set [5], or Chinese "Tao" or something another like in philosophy or religion. And it is rather possible, e.g., that a highly experienced Buddhist will say after reading this article something as: "all in the article is a nonsense, besides what is about Zero element. In reality – there was Nothing, there is Nothing and there always will be Nothing, Shunia, Emptiness". But a not too experienced Buddhist will go away from a rail if on this rail a tram moves, even at that somebody tells to him that the tram is an illusion; recognising by such a way the existence of an objective reality.

So do the authors, but here we try to show that the Information concept can be useful outside philosophy or religion also, in this case – in Physics.

Albeit it's rather probable that Nature sciences, including Physics, will eventually explore in future the Set as whole, now its study only so called "matter phenomena". Because of in philosophy and so – in Nature sciences - there isn't of a consensus in the problem – "what is material/ non- material?" in [1] following *criterion for the Matter* was suggested: a *process/ phenomenon/ entity is material if it exchanges* (interacts) with other process/ phenomenon/ entity *exclusively by true information*.

If a process/ phenomenon/ entity has a capability to produce/ apprehend false information, then it isn't material in some sense; some examples: "quasi- material" - living beings; "non - material" – a human's consciousness (at least – at the information processing), religious phenomena. All these examples relate to distinct, only partly (non-overlapping at all?) overlapping, subsets of the Set when now just the elements (and, of course, its' interactions) of the subset "Matter" are studied by physics. A number of existent non-material subsets and the subset "Matter" constitute the subset "our Universe"

Another important property of the information is that the *information can be* (any information always is?) *"absolutely accurate"*. But since (i) - any element in the Set is always connected informatively with all "utmost infinite number" of other elements of the Set, including "in any times" of it's own existence (as well as of it's "non- existence"), and (ii) - just because of this accuracy a little change in a formal language representation of some information can drastically change the context, *there exist such a phenomena as the randomness and the bifurcation.* E.g. – possibly the chaotic connections of a particle with all Set's elements lead to quantum-mechanical uncertainty of the particle's parameters. Though just because of such an uncertainty material systems constituting of more then one material objects can change in the way as we observe, what was shown by Zeno yet 2500 years ago; in fact Zeno predicted, if not the QM as a whole, but at least the Uncertainty principle.

So in the Matter the particles/objects/, systems of objects/particles, exchange by only logically true informational "messages", i.e. the subset "Matter" (including our material Universe as a whole) is some like as a computer. Such a idea isn't, of course, new - hypotheses that our Universe is a large computer appeared practically at once with the appearance of usual computers (see, e.g., [6 – 17], though the list can be much more), but that were only the hypotheses which had not necessary grounds (if, of course, one doesn't consider a Creation of Universe as of a logical structure from nothing by some omnipotent judicious Being, Who "established the laws"). Now this idea becomes be grounded, moreover – the absence of logical structure of the Matter (what realizes as "Nature laws" in the Universe) would be rather surprising.

Note here also a couple of another sequences of informational concept: (i) - since the Information elements can exist only as a number of logical connections and realizes as a choice of some alternatives, the Information Set must be "countable" (discrete) Set. So, albeit the Set's cardinality is utmost maximal, any "realization" of possible relatively independent subset, e.g. – of our (at least – of it's material subset) Universe, should be discrete, and (ii) - since a language exists on the Zero element and since the Zero element coincides with the (total) Information set - hence the contain of any subset of the Information[*)] can be expressed by using some language.

## 2. SOME CONSEQUENCES FROM THE INFORMATION CONCEPT – A PHYSICAL MODEL

### *2.1. The Space and the Time*

Some informational system, which contains more then one element, should contain also some logical condition(s) to be existent just as a system of elements, i.e. – a condition(s) that should realize in this system informational differences between the elements. In the system "our Universe" such utmost common conditions are the Space and the Time, which are, naturally, some informational systems (codes, rules) also. As it was noted above, since the information can be always reduced to the choice of possible variants, any realization of the information - all in the World - must be something discrete, so Space and Time are some discrete systems.

---

[*)] Including any subsets of false information

Many authors [9], [10], [15],[16], [17], etc., point out that the Matter in our Universe is some rather logically simple system (in the "Universe computer" rather simple program code runs). That follows from the fact that the number of "Nature laws" is not large, when laws themselves are rather simple and can be reduced to a number of the groups of high-level symmetry.

It seems rather evident, that to exist as some stable isolated system/ subset (e.g., - the Matter) under incessant impacts of the rest elements in the Set, it is necessary for this system to be made from relatively stable ("*fundamental*") logical elements, i.e. *from closed logical systems where logical bonds are much stronger then these impacts*.

We don't know now – what are the logical structures of these elements, though some common reasonable suggestions for our Matter were made (see, e.g. [8], [10] and Refs. in these articles). First of all, for the Matter to exist as some "independent" system in the Set, it seems necessary for any subsystem of the Matter and for the Matter as a whole to be constituted from some *logical elements*, which should be *reversible*. Then the system doesn't dissipate the energy (some physical variable, see below) and so requires no additional energy to exist and to change/ develop. Besides in physics were obtained some values for fundamental quantities in the Universe – Planck units. And these units aren't change near a century already, regardless to the fact that the physics went far ahead for this time; what indicates that these units are indeed fundamental. So for *Space a fundamental unity appears – Planck length*, $l_P$, $l_P = (\frac{\hbar G}{c^3})^{1/2}$ ($\hbar$ is reduced Planck (Dirac) constant, $G$ - gravitational constant, c- speed of light) - and this length is, as it seems (more below), *the size of fundamental, at least be-stable, logical elements (FLE) which are used to build the Universe's Matter*.

About the Time there is, e.g., well known "definition" of J.A. Wheeler: "Time is what prevents everything from happening at once." That is a joke to some extent (and note – really in the Set "everything is happening at once always" principally), but it becomes indeed correct if will be as "Time is what prevents cause-effect events from happening at once."

So *any of the elements,* at least - in the Matter, i.e., - elementary particles, systems of particles, etc.- *have own (proper) times*, but, *since they* are eventually *constituted from the same* FLEs, *there is the fundamental* (and universal) *unity of time which is the time interval need to change the state of the FLE* (to flip the FLE). At that, since Matter code is simple and highly "standardized" for the particles/ objects in Matter, – the time can be used (and is used) as some global/ universal variable to describe physical processes. The fundamental unity of time is Planck time, $\tau_P = l_P / c \approx 5.4 \cdot 10^{-44}$ s, where $c$ is the speed of light. In this formulae Planck time is defined as some derived unit (through the fundamental Planck length and "fundamental" speed of light) but *really the fundamentals are Planck time and Planck length, when the speed of light should be defined as derived unit.*

Both – Time and Space - have analogous features: both are utmost common conditions/ rules for the systems of elements to exist in the informational structure "Matter", both have fundamental (minimal) intervals to separate the elements in the Matter; perhaps main distinctions are that (i) - Time "quantizes" only logical sequences of events, when Space quantizes logically distinct fixed information, and (ii) - a human can observe directly by senses only fixed (i.e. - spatial) information and so doesn't "see time". From this it follows that rather probably Time and Space as informational systems have analogous structures – in "traditional" physics that realizes as an equivalence of the space-time coordinates; when in informational physics that indicate also on some specific logical features of FLEs.

At last note here, that: (i) - for some precaution in [1] (and here below) it was suggested that in the Matter there exist at least two kinds of be- stable FLEs, namely - the FLEs which constitute some informational structures (IS), i.e. material objects (e.g., – elementary particles), "t-FLEs"; and the Space FLEs ("ether FLEs") – "e-FLE s"; and (ii) the consideration above has in some moments the analogues, e.g. - to the presentation of the Space as a "spin network" (in this concept – as a "FLE network") by R. Penrose [7]. *Though it is very probable that t-FLEs and e-FLEs have the same* (practically the same) *structures to make possible the movement of the particles in the Space, as well as to make possible the creation of new particles at particles interactions.*

## *2.2. The "development" (the realisation) of the Information in material World*

### *2.2.1. Elementary (subatomic) particles*

So the informational approach means that anything in the Universe (at least in the Matter) is/are transformations (under exchange by information) of the informational structures (IS) and that the elementary particles are some primary ISs also. Correspondingly in [1] were suggested for a particle two options of "informational currents" (IC) – "time IC" and "space IC", and one option for fixed information using only some common physical parameters and Dirac constant (elementary action), $\hbar$:

- the time IC (t-IC):

$$j_t = \frac{1}{\hbar}\gamma m_0 c^2, \qquad (1)$$

- the space IC (s-IC):

$$j_x = \frac{1}{\hbar}\gamma m_0 c^2 \beta^2, \qquad (2)$$

- fixed information:

$$\Delta I_M = \frac{\Delta M}{\hbar}. \qquad (3)$$

($c$ is the speed of light in vacuum; $\beta = v/c, \gamma = 1/(1-\beta^2)^{1/2}$ is the Lorentz – factor of a particle movement, $\Delta M$ is angular momentum, $m_0$ is the rest mass. The dimensionality of the time and the space currents is [bit/s], the dimensionality of fixed information is [bit]). Besides note, that fixed information relates, quite naturally, also to the physical action, $S$.

Though it should be noted that at a movement of a particle the uncertainty of its parameters is defined by the Uncertainty principle and minimal increment of the parameters (at least of the action) that corresponds to changing of information on 1 bit is equal the half of Dirac constant: $\Delta S = \Delta x \cdot \Delta p = \hbar/2$. That can require of some modifying of the physical model presented in next sections of this paper, but at least in first approximation the corrections rather possible are as not too essential and further we adopt the correspondence "an increment of action (or angular momentum) – an increment in 1 bit of information" be equal $1\hbar$.

An elementary particle in this model can be roughly represented as some closed loop linear structure of t-FLEs which are continuously flip (so – its are closed loop "FLE currents"), when "universally significant" (further – "us-FLE") to the external Matter are the FLEs, which flip in the end of FLE-line having length be equal Compton length of the particle, $\lambda \equiv \frac{\hbar}{\gamma m_0 c}$; the radius of this (circle) loop is equal to the Compton length, $r = \lambda$. The rate of us-FLE flips is the time IC in Eq. (1). For such a conjecture there is a number of reasons, e.g. as: for static condition in a particle "an active point" of flipping FLE moves on the loop with speed of light. Then the energy of this point is $E = pc$, when momentum is equal $p = M/\lambda$. For a particle having the *"point's angular momentum"* equal to *particle's spin*, e.g., ½, $p = \dfrac{m_0 c}{2}$ the energy of the "point" is $E = m_0 c^2 / 2$ - i.e. the value, which is not too far from the real one.

From the fact that the stable particles' currents never stop, follows that the t-FLSs, as well as the loop cycle codes defining the characteristics of particles, are reversible codes. So it seems rather reasonable to suggest that, e.g. – *"direct" codes relate to the particles when reversed ones – to antiparticles*. At that the antiparticles "live" as if in opposite time direction.

Note here also a rather trivial but important corollary of this informational concept for the atomic/ subatomic objects/ particles – now becomes clear one of main QM postulates – *the postulate of identity of the same kind particles*: the information is unique thing that can have absolutely identical copies, so do the particles as some informational clones.

Since FLEs in a particle sequentially flip, particle's FLEs in statics "remain on its' Space places" but move in Time, when through Space "virtual flipping point" moves. If a particle moves in Space as a whole, then additional degree of freedom appears – sometimes particle's FLEs must execute "Space flips" (possibly as exchange of the particle's t-FLE by e-FLE if its are identical-?) to change particle's Space coordinates. Any space flip require some time, and it is rather probable that this time is equal to Planck time also – in fact any space flip is simultaneously the time flip of the FLE – and eventually of the particle - in Space and Time. But with fundamental limitation – resulting flips rate cannot exceed in any space-time direction the inverse value of Planck time.

***Algorithmic model for a material particle in statics.*** So, at least for the statics, a particle in the space-time is some circular dynamical object that always moves in static conditions only in the time direction having some variable that we can call "momentum", *p₀*, which is equal $\vec{p}_0 = m_0 c \vec{i}_t$, i.e. directed along *t*-axes; at that FLE flipping point moves in the space through the circle with the radius, $\lambda_{t0}, \lambda_{t0} = \hbar / p_0$ and *m₀* is some coefficient (Fig.1). The magnitude of the flipping point's momentum is equal also to *p₀*; corresponding angular momentum is equal to elementary action, $\hbar$.

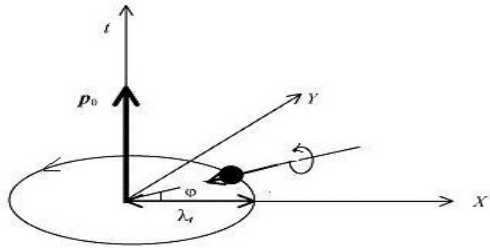

Fig. 1. A particle in statics. Large black point in the circle is flipping FLE. The movement of a particle as of a *singled out specific informational structure* along *t*-axis is step-by step and the step's length is equal to the circle radius, $\lambda_{t0}$.

***Algorithmic model for a material particle at a movement.*** From above follows a number of rather reasonable conjectures.

(1) So after non-zero impact on a FLE in a direction this FLE starts a flipping – and a movement of the flipping point - in two, i.e. in Space and Time, directions. At that it is inessential – this FLE was in statics or it was flipping on a *straight* line of sequentially flipping (with speed of light) FLEs. Indeed, in last case Compton length of corresponding "particle" is infinite, so the particle has zero inertia (zero "rest mass") independently on it's being at rest or in a movement after *infinitesimal* impact on the FLE. And the alternative flipping occurs because of the FLE cannot to flip in initial direction with a time that is lesser then Planck time and so it is forced to flip in another Space / Time direction. Thus a non - zero impact on a FLE means occurrence of an additional specific information in space-time that reveals itself as a specific particle.

At that, as the experiments show, only a number of possible stable – and different - algorithms can be realized in the FLE circle, from what follows that FLEs have more sophisticated structure then simple bi-stable one. The clearing of this structure is the task of the future, but to make more understandable the foundations of some existent theories – special relativity and quantum mechanics - seems be sufficient to take into account only bi - stable FLEs.

If the impact - and corresponding momentum – is directed along time direction (along of t - axis), then "usual material" article (further – "M-particle") occurs. An impact in space directions results in occurrence of another sorts of particles. Since the transformation of initial FLE straight line into a "helix" doesn't change the direction of movement (after the impact – of the particle's movement), which remains be, in certain sense, uniform and rectilinear, M- particles obtain an inertia (the mass) when move (next section) in Space, when in Time they remain be "restmassless". A particle, which appeared after "space directed" impact, obtain the mass in Time, when in Space it remains be "restmassless". For example electrons, protons, etc. appeared after "time directed" impacts, when to produce a "restmassless in Space" photon is necessary to act on a flipping on a straight *space line* FLE in a space direction.

(2) The model above leads to following natural conjecture: since at a M-particle's movement its flipping point moves always with the speed of light, then any additional impact in a Space direction results in occurrence of next "helix". At that (i) - corresponding momentum is proportional to the speed of the particle, $\vec{p}_x = \mu \vec{v}$, and (ii) - the impact leads to a movement of the particle *as a whole* in two directions – along the speed vector and to circular movement in the plain which is perpendicular to the speed vector when it the radius of this circle is equal $\lambda_s = \hbar / p$ - i.e. is equal to de Broglie wave length value. An example when the momentum is directed along X-axes is shown in Fig.2.

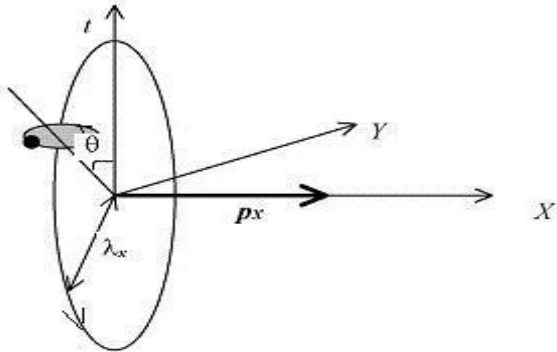

Fig.2. A particle's movement along *X* axis as a combination of two circular and one direct motion. Big black point in the lesser circle is flipping FLE. Non – relativistic case, so momentum $p_0$ is too large to be shown in the figure.

Resulting momentum (see Fig. 3) is equal: $\vec{p}_r = \vec{p}_0 + \vec{p}_x$ and, since any space impact is always perpendicular to t-axis and so momentum $p_0$, $\vec{p}_0 = m_0 c$ is "relativistic invariant", module of this momentum is equal

$$p_r = (p_0^2 + p_x^2)^{1/2}, \qquad (4)$$

when "space-time step" in $p_r$ direction is equal to corresponding wave length, $\lambda_r = \hbar / p_r$.
*Since FLE flipping time and FLE size are constant*, for one second the particle moves along $p_r$ direction on the distance that is equal to the speed of light, $c$, when in the X-direction – on the distance $v = \beta c$.

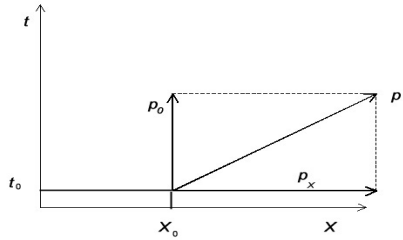

Fig.3. A momentum, $p_r$, of a M-particle after space – directed impact with momentum $p_X$.

Correspondingly we obtain for the momentum $p_r$ another equation:

$$p_r = \frac{p_0}{(1-\beta^2)^{1/2}}, \qquad (5a)$$

and for $p_X$:

$$p_X = p_r \beta = \frac{m_0 v}{(1-\beta^2)^{1/2}} \equiv \gamma m_0 v \qquad (5b)$$

So $\mu = \gamma m_0$, $p = \gamma m_0 v$ - as it is used in the momentum's definition in current SR theory.

From above follows that one "flip" of a FLE corresponds, in certain sense, the rotating of the FLE on angle equal to 1 radian. That lets to introduce the variables "rate of rotation" of the FLE, $\omega$, and "moment of inertia" of the FLE, $J$. The rate of rotation is vector that is perpendicular to momentum and magnitude of $\omega$ is equal to the value of corresponding informational current. On first sight one can expect that $\omega_r$ is the sum of $\omega_0$ and $\omega_X$, but it is not so. The projections of $\omega_r$ on $t$ and $X$ axes are $\omega_{rt} = \omega_0$ and $\omega_{rX} = \gamma \beta \omega_0$, when $\omega_X = \frac{v}{\lambda_X} = \frac{v p_X}{\hbar} = \gamma \beta^2 \omega_0$.

Such a situation arises again owing to the fundamental limitation on the flipping rate that was pointed above. At that, since an impact in Space doesn't change the momentum in t-direction, the t-step doesn't change also and so actual t-rate value is $\omega_t = \omega_0 (1-\beta^2)^{1/2} = \omega_{rt}(1-\beta^2)$.

From above follows a couple of rather important implications.

(i) Since actual flipping rate for moving particle becomes be slowed down, it means that particle's algorithm becomes be slowed down also. If the particle isn't stable and there is a probability of a "soft failure" on some loop tact (and the particle decays) the slowing of the rate leads to that the (half-) life of such particle increases comparing to the case when the particle is at rest. And – if a system consists of particles - all/ any processes in this system become be slowed down on the factor $1/\gamma$ also. For example moving clocks – mechanical, electronic, biological – will show lesser time then at rest.

(ii) . On first sight, since Compton length of moving particle decreases, the particle's dimension should be lesser then when the particle was at rest. But, since "space" impact doesn't affect on the particle's "time Compton length", the particle's dimension in space-time remains be the same as it was when particle was at rest, simply trajectory of flipping FLEs becomes be mach more complicated comparing with a circle. But initial circle becomes be rotated on the angle $\varphi$, (see Fig. 3) $Cos(\varphi) = (1-\beta^2)^{1/2}$. So *the projection of moving particle on a Space* plain becomes be shorter in the direction along the movement. In SR theory this effect is known as "Lorentz contraction".

***Informational approach and the basis of the SRT.*** All equations above well correspond with their analogues in special relativity theory. But there is rather important difference. SRT considers "reference frames" which move relative to each other. At that coordinates (*t*, *r*) of any point (and the point itself) in the spacetime can belong to any of the frames, when the relation between coordinates are determined by Lorentz transformations. This leads sometimes to that in some books, where the SRT is described, one can meet the expressions something like as "at a movement the space transforms into the time and vice versa". Though that seems evidently incorrect – for example it seems as not too plausible to think that, e.g., every electron in an accelerator transforms space-time in whole our Universe.

And that indeed isn't so. Since any particle (or a system of interacting particles) has its *own specific time and space parameters in the absolute Space ad Time of Universe*, a movement of the particle changes only on the parameters of this particle (system of particles) and nothing does with the external World.

As an example let us consider a classic SRT task (see Fig. 4): on a platform, which is perpendicular to *X*-axis, there is a source of the light (in point *A*) and the mirror (in point *B*). In point *A* there is also a moving clock which measure time that is spent for the light's pass to the mirror and back (pass $A_1$-$B_1$-$A_1$ on Fig. 4).

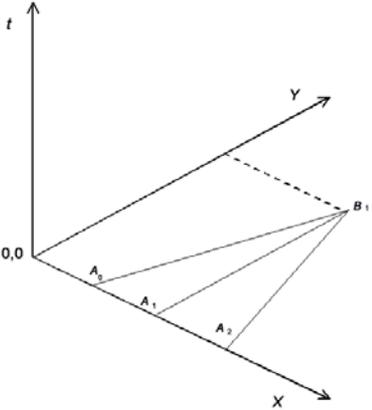

Fig. 4. The light path in space if the light source moves with a speed *v* along *X*-axis.

The platform (reference frame *K'*) moves in some reference frame *K* with speed *v* along *X*-axis.

Here we have the case that is rather similar to the case when M-particle moves in a frame, which was considered above (Fig. 3), but the cases are different. The M-particle moves in (*t, X*) plane, when light (the photons, see above) moves in certain direction in *space* – (*X, Y*)- plane; moving there and back along *t*-axis. Since speed of photons in reference frame *K* is equal *c*, from Fig. 4 immediately follows that *real* speed of light in the frame *K'*, *c'*, *is not equal to actual speed of light, c*, $c' = c(1-\beta^2)^{1/2}$ (so here a "*space dilation*" take place, which is, generally speaking, independent on time dilation). But since moving clock shows the time dilated on the same factor, the *measured* speed of light in the frame *K'* is equal to *c* also.

Let us consider another standard SRT task (see Fig. 5): a wagon having the length $l_0$ moves along *X*-axis with the speed *v*.

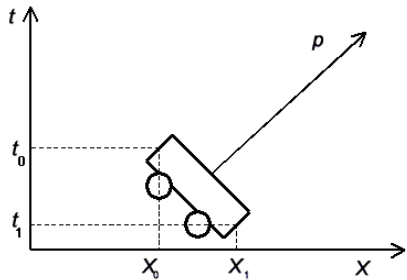

Fig. 5. A wagon having the length $l_0$ moves along *X*-axis with the speed *v*.

From Fig. 5 evidently follows the first equation of Lorentz transformation:

$$x = vt + x'(1-\beta^2)^{1/2}, \qquad (6a)$$

and (though after more complicated work) the second equation:

$$t' = (1-\beta^2)^{1/2} t - \frac{vx'}{c^2}, \qquad (6b)$$

but with essential correction – these equation are valid not in whole Space but *are true inside of the wagon only*: $x' \in (0, l_0)$, $x \in (x_0, x_1)$; $x_0 = vt$ and $t' \in (t_0, t_1)$; $t_0 = t(1-\beta^2)^{1/2}$

The *t*- decrement along the wagon's length appears at the acceleration of the wagon up to the speed *v* and further remains constant at the uniform movement. So if one synchronizes a couple of clocks in the ends of the wagon before the acceleration, then he always can measure the wagon's speed relative to the start reference frame. And if somebody could make such a synchronization (and a wagon, of course) just before Beginning of our Universe, then we were capable now to measure our speed relating to absolute (Universe's) reference frame.

But, regrettably, it seems that such experimenters weren't made in that times and what we can do now – that only measure this decrement relating to our own reference frame. From Eq.6 follows, that if a wagon has the length 100m and if the wagon is accelerated up to speed 1000m/s then the decrement will be ~$5 \cdot 10^{-13}$s – the value that, rather probably, now can be measured.

***Informational approach and the SRT.*** So we should conclude that the first – Lorentz's – version of the [Aether] theory was true in the points of absolute Space, Time and Aether, though in all other aspects standard version of SRT is mighty and convenient mathematical tool, which allows solving seems any practical problems in mechanics and electrodynamics.

As well as Eqs. (1)-(3) can be quite naturally expanded into 4-dimensional representation, then the 4-time current becomes as (let here to use symbols without attention to covariance/ contravariance):

$$j_\mu = (\frac{E}{\hbar}, \frac{c\vec{p}}{\hbar}) = (j_t, j_t\vec{\beta}). \qquad (7)$$

If some point in a reference frame has the coordinate $x_\mu = (t, \frac{\vec{r}}{c})$ then scalar product

$$\Delta I = j_\mu x_\mu = \frac{Et}{\hbar} - \frac{(\vec{p}\cdot\vec{r})}{\hbar} = \frac{1}{\hbar}[Et - (\vec{p}\cdot\vec{r})] \qquad (8)$$

formally turns out to be the quantity of the information (the number of us-FLEs' time flips) for a particle needed to reach this point, if the particle has coordinates (0,0). Let us here don't consider the realizability of such a movement (there is the problem of the compatibility of *t, x* values when taking into account finite speed value of the interaction/ information propagation), note only, that the expression in Eq.(8) is the part of exponent index in the [QM] wave function of free particle.

If we use in Eq.(8) the parameters of the particle's actual movement then obtain:

$$\Delta I_s = j_{s\mu}x_{s\mu} = \frac{1}{\hbar}[Et - (\vec{p}\cdot\vec{r})] = [\vec{r} = \vec{v}t]$$
$$= j_t(1-\beta^2)t = (j_t - j_x)t \qquad (9)$$

From Eq. (9) it follows that the rate of the information change at a particle's movement is equal;

$$\frac{dI}{dt} = j_t(1-\beta^2) = \frac{1}{h_D}m_0c^2(1-\beta^2)^{1/2}. \qquad (10)$$

From Eq.(10) follows that $\frac{dI}{dt} = -L$, where *L* is Lagrangian for free particle.

From Eq. 9 follows that the change of information on some trajectory is, in fact, physical action. Note here, that on the correspondence "the change of information on some trajectory – physical action", as it seems, was firstly pointed out in [9].

The informational currents $j_t$, $j_x$, correspond to the FLEs rotation when the turning's angle of the angular momentum of the flipping FLE in the plane that is perpendicular to $\vec{p}_r$ is the sum of the turning's angles in the in (*X,Y*), i.e. – the angle $\varphi$ (see Fig. 1), and (*t,Y*), i.e - the angle $\theta$, (see Fig. 2) planes.

The $p_x$ projection on $p_r$ "is responsible" for the "Space part" of resulting angle which is equal $\Delta j_r t = j_r\beta^2 t$, when the rest of $j_r$ is responsible for the "Time part" of resulting angle. The "Time part" phase $\Phi = j_t t - j_x t$ at the particle's movement corresponds to the particle's physical action Eq. (9).

A particle "obtain" a specific position relating to external Matter only when it's us-FLE flips, or after the angle's $\varphi$ increment becomes be equal 1 radian. Between these moments the position (and possibly some other properties of the particle) are uncertain for the external – analogously if in a computer a program code runs, the state of the code becomes uncertain on the time need for some electronic gate to flip. Moreover, if a code contains some subroutines – the state of the code becomes uncertain on the time need for next subroutine to carry out its calculations.

*So from this consideration follows direct correspondence of Quantum Mechanics and Special Relativity theories, when de Broglie wave is the projection of "$j_t$ wave" on a Space plane.*[*)]

### 2.2.2. Mediation of the forces and complex systems

Now four "fundamental" kinds of the interactions (four "fundamental" forces) are known – gravitational, weak, electromagnetic (EM), strong; which differ, e.g., for the proton as (approximately) $10^{-36}:10^{-11}:1:10^{3}$.

However note what, generally speaking, seems evident – some other physics laws should be a result of corresponding forces impacts, besides those that are a corollary of intrinsic information properties, as it is in the case of particles identity postulate above. E.g. – some force must impact on the electrons in an atom's shell to hold their spins under Pauli exclusion principle. Such forces aren't included into "fundamental forces set" in physics now and are called in traditional theory as "purely quantum mechanical effects". But "to call" in some formal theory doesn't mean to have understood - why does that work? Here we can say that these "purely QM effects" are in reality some "first level" logical laws in informational physics. In this physics any "physical" force relates to corresponding exchange by logic messages between the elements in a specific IS, e.g. – in the case of Pauli principle - between the electrons in the atom, and so a way appears – how to make the nature of things more clear.

Since above the correspondence "rate of us-FLEs' flips – the energy" was already introduced for a particle, it seems rather logical to expand it on more complex systems, e.g. – atoms. For example – well known N. Bohr postulates reduce to the statement that some change in an atom the electron's angular momentum state can be carried out only as the change in the information on integer number of bits.

Earlier in the model [1] it was suggested that any force has special mediating particle which appears as some intermediate algorithm aimed to synchronize and to unify in some part (on some time) the steps of t-ICs of interactive particles, if it is possible. Then the rate of unified steps of particles is the potential energy, $U$, of the system, when at each unified step interactive particles get (very roughly!) the momentum $p_0 = \pm \hbar / r$. But it seems such a suggestion meets with the difficulty when relates to the fact of finite speed of a force transmission ($\leq c$). So more reasonable seems the suggestion that at an interaction of a mediator with a particle *some t-IC step* only in this particle *becomes "be spent" by interaction*, resulting in the t-IC's decrease (if $U<0$; $p_0 \propto -\vec{r}/r$) and in corresponding mass defect; or "*be added*" resulting in t-IC increase if $U>0$, $p_0 \propto \vec{r}/r$.

More specific consideration of complex systems requires additional work and now is out the scope of this article. But a common notion can be once again mentioned. As it was pointed out above, one of main problem of the informational concept is the problem of material systems stability in the Set' *infinite (and always active)* environment – material informational systems should be closed logical systems where logical bonds are much stronger then environment's impacts. Now we can add, that as a rule utmost stable in the Matter appear to be the cyclic systems and the Matter, as it seems, seek eventually to build such a systems – particles, atoms, planet systems, etc.

### 3. SOME CONSEQUENCES FROM THE MODEL - GRAVITY AND ELECTRICITY

There are a number of theoretical gravity concepts now that appeared since famous Newton work. The main drawback of the Newton theory is that it is true in a "static" (or non-relativistic) case only. When

---

[*)] And we can conclude that Zeno predicted not only the QM, but SR theory also.

a body becomes to move, Newton's law becomes be invalid. To overcome the problem of gravity forces at a movement there were developed a number of theories, which are based on the analogy of Coulomb and Newton laws and, further, on the corresponding analogy with Maxwell's theory of EM force. The first attempt, probably, was the O. Heaviside work [20], another – and utmost known – is the General Relativity theory.

All these works contain the main – and, as it seems, unresolved till now problem - "What is analogue of E-M magnetic force?" In GR so called "gravitomagnetic" force is introduced, but the analogy with the EM exists only for week field approximation. Such a situation leads to some strange conjectures, e.g. - there are some articles where it is shown, that GR concept leads to conjecture that there exists some "critical" body's speed when gravity force changes the sign: "…This equation contains a critical speed $v_c = 1/(3)^{1/2}$; that is, for motion with $v < v_c$, we have the standard attractive force of gravity familiar from Newtonian physics, while for $v = v_c$, the particle experiences no force and for $v > v_c$ the gravitational attraction turns to repulsion. These results are valid in the linear approximation for the gravitational field…" [21]. Another example – the work [23], where so called "Gyron field" conjectured.

To test the Informational concept in application to the gravity and electricity theories in [1], v5, a well known standard problem (e.g.[22]) of the forces between two moving bodies having charges and masses was considered, as, e.g., that was made in [23], [24].

### 3.1. The gravity: static solution

Remaining in the informational concept it is possible to put forward [1], rather reasonable conjecture that it seems - since the gravity force is universal (regardless to the kind of particles) and acts always as an attraction - there is only one case to satisfy to this condition: if the gravitational potential energy of a system of some bodies *is proportional to **the accidental coincidence rate** of some equivalent of the t-ICs of the particles* of these bodies *that always exists if the t-FLE's flip- time is not equal zero; when in gravity interaction only us-FLEs, i. e. the FLEs that are used for movement/ localization of particle in Space, "take part"*. From this suggestion it follows that (i) – any particles interact trough the gravity, (ii) - the gravitational force between some bodies must be a result of the interactions of smallest (that aren't divided on a components in real conditions) particles that constitute the bodies; and (iii) – the gravity force should be very weak. The problem for complex (many particles-) bodies was considered in [1]; here, to simplify the equations, only "two - particle" gravity case is discussed.

For two particles at rest having gravitational masses $m_1$, $m_2$, placed on the distance between the particles, $r$, the "Newtonian" potential energy is

$$E_{gN} = G\frac{m_1 m_2}{r}, \qquad (11)$$

where $G$ is Newtonian constant of gravitation.

Basing only on approach of section 2 above, it can be shown that, under rather plausible assumptions, the potential gravity energy can be expressed also as follows.

As to section 2, the t- and e-FLE's sizes are equal to Planck's length, $l_P$. and
- assume that – since the gravity is universal force, so it doesn't depend on any additional information that differs the particles - hence in gravity interaction only us- FLEs (i.e. – single FLEs) "are used".

Assume, also, that at every t-IC step of a particle in Space a "rim" of e-FLE's flips starts expand with radial speed be equal to the speed of light, $c$, so the rim's area is equal $2\pi r l_P$ ($2\pi c t l_P$); and

- the times, $\tau_i$, of the t-FLE's flip and of the interaction of the e-FLEs and t-FLEs, $\tau_r$, are the same and are equal $\tau_i = \tau_r \equiv \tau = l_P/c$, i.e. – to Planck time.

It is well known – at least for any nuclear experimentalist – that if random pulses having lengths $\tau_1$ and $\tau_2$ come with rates $n_1$ and $n_2$ in the inputs of a coincidence circuit, then the rate of accidental coincidences is equal $n_{cc} \approx n_1 n_2 (\tau_1 + \tau_2) = [\tau_1 = \tau_2 = \tau] = 2 n_1 n_2 \tau$. So the accidental coincidence rate in the particle 2 when radiates the informational current of the particle 1, $N_{cc21}$, is equal:

$$N_{cc21} \approx \frac{j_{t1} \cdot 2\pi r l_P}{4\pi r^2} \cdot P \cdot j_{t2} \cdot 2\tau = \frac{m_1 c^2 \cdot 2\pi r l_P}{4\pi r^2 \hbar} P \frac{m_2 c^2}{\hbar} 2\tau \;; \qquad (12)$$

where $P$ – is the probability of particles' t-FLEs interaction if a rim of the particle's 1 t-IC e-FLE flips passes through the t-FLE of the particle 2.

Since the system is symmetrical, the coincidence rate of both bodies is equal $N_{cc21}$ ($\equiv$ "gravitational coincidence current", $j_G$) and the potential (binding) gravitational energy is equal ($P=1$):

$$E_{gI} = \hbar \cdot N_{cc21} = \frac{c^3}{\hbar r} l_P^2 m_1 m_2$$
$$= \frac{l_P^5}{\hbar r \tau^3} m_1 m_2 \qquad (13)$$

where $\vec{p}_0 \equiv -\hbar \vec{r}/r^2$ and $m_1, m_2$ are inertial masses.

From Eq.(13) follows that $E_{gI}=E_{gN}$.

From Eqs.(1), (12), (13) it follows, also, that - at least for static conditions - *inertial and gravity masses are equivalent – simply both variables (as well as – the energy), are "created by" the same informational time current.*

*Besides note, that here we don't use any experimental data on the strength of gravity force – only some rather reasonable, and, besides, – which follow from earlier common conjectures about the Matter's objects structure – suggestions.*

And so we can conclude that not the gravity constant, $G$, but Planck length, Planck time and elementary action are indeed fundamental. As a support for such a conclusion and, in addition, – as a ground that Matter is a program code operating with rational numbers, - we can add that if somebody uses these fundamental constants instead of the constants $G$ and $c$ in the equations for derived Planck units (Planck mass, Planck momentum, etc.) derived units become be rational equations, which constitute complete set of the values to describe Nature. Though there is one exclusion – Planck charge, but it probably that it is rational value also - the fine structure constant, $\alpha$, is probably equal to rational fraction, e.g., $\alpha \approx (21354/249975)^2$; when, as it seems, Dirac constant and speed of light are "perfect square" values. Note also, that at least for the static conditions the rims of a particle transmit at gravity interaction to any another particle all information about the localisation of the first one in the vector value of elemental momentum $\vec{p}_0$.

Though with practically 100% QM uncertainty of the distance.

It is evident that from Eq.(12) follows equation for the gravity force in the statics:

$$f_{G0} = \frac{dP}{dt} = N_{cc21} p_0 = \frac{Gm_1 m_2}{r^2} \quad (14)$$

It can be noted here, that from Eq.(12) it follows also, as it seems, some interesting result. If there is a couple of bodies having masses $M$ and $m$, $M>>m$, then the fraction, $p$, of the gravitational coincidence t-IC is equal

$$p \approx \frac{j_G}{j_m} = (\frac{GMm}{r\hbar})/(\frac{mc^2}{\hbar}) = \frac{GM}{rc^2} \quad (15)$$

Since increments of t-ICs in both bodies are equal, for given body the $j_G$ fraction is equal ½ p. As it was assumed earlier when in an IS, e.g., in an atom, a "gravitational coincidence" happens, then corresponding IS's t-IC step is not used in own IS's algorithm. So for this IS it's "own" t-IC becomes slowed down and the IS becomes "to live in slowed time", $t_{own}$ that is evidently inversely proportional t-IC and is equal in first approximation to:

$$t_{own} \approx t_0 (1 - \frac{GM}{2rc^2}) \quad (16)$$

where $t_0$ is the IS's time for free body. This value is, in turn, in a first approximation, equal to the gravitationally dilated time in GR:

$$t_{GR} \approx t_0 (1 - \frac{2GM}{rc^2})^{1/2} \approx t_0 (1 - \frac{GM}{rc^2}) \quad (17)$$

The difference is in the denominator (factor 2) in ratio (17), but that is usual discrepancy of Newtonian and GR gravity theories – "Newtonian" angle of light declination in gravity field is ½ of the GR one, the radius of "Newtonian" black hole is twice lesser then for GR, etc. This consideration till now corresponds, in fact, to the Newtonian model and should be, of course, additionally further developed to take into account GR approach.

### 3.1.1. Experiment?

From Eq.(12) follows also that the value of the gravitational (coincidence rate) current in any particle is random in the time. Correspondingly so does the gravitational force that impacts on this particle and some uncertainty should appear at gravitational interaction of small masses. The detection of corresponding randomness of the gravity force (or some equivalent physical value) will be rather weighty evidence that suggested informational concept is true. Possible experiments were suggested in [1], [2], [18], [19], but in these Refs. Planck's constant $h = 2\pi\hbar$ was used, so the elemental momentum value, $p_0$, was, rather probably, $2\pi$ times more then real value. Correspondingly, gravity randomness' impact in these Refs. was overestimated and it is necessary to repeat the calculations. In addition – some analysis of the impact values obtained in the Refs. shows that at decreasing the momentum value the majority of considered experiments – e.g., with the atoms – rather probably become outside the capabilities of existent techniques to detect the effect (though it can be interesting to consider an experiment with interference of ultra-cold neutrons), besides the experiment with measurement random distortion of monochromatic photon beams in maser band. Below some results for the photon beam case are presented.

So (see, Eq.(2) ) common gravitational t-IC of two bodies having masses $m_1$, $m_2$ is equal:

$$j_{gN} = G \frac{m_1 m_2}{\hbar r} \qquad (18)$$

and the gravity force is quantified and it should be stochastic if some small masses interact.
If the body 2 is a photon with frequency $\nu_0$, the mean current (further - $n_\gamma$) is equal

$$n_\gamma = 2\pi G \frac{m_1 \nu_0}{r c^2} \qquad , \qquad (19)$$

and at every gravitational interaction photon's frequency becomes shifted on the value $\Delta \nu = c / 2\pi r$.
The time for the photon to fly through the distance $l$ is equal $l/c$, so the number of the gravitational interactions of a photon in this distance is a random number that is distributed under Poisson law:

$$p(k) = \frac{N^k e^{-N}}{k!} \qquad (20)$$

with the mean value $N = n_\gamma l/c$. In zero approximation [2], in near surface Earth gravitational field, $r = R_E$, $N \sim 8.8 \cdot 10^{-17} \nu_0 l$ and the frequency shift value, $\Delta_E \nu \approx c / R_E \equiv \Delta f = 7.5$ Hz. So after the path $l$ (e.g. upward directed) initial beam should be split on a number of modes having frequencies $\nu_0$, $\nu_0 - \Delta f, \ldots \nu_0 - k\Delta f \ldots$ having the intensities that are proportional $p(k)$.

However, Earth is a rather large body and the limits for $r$ are rather large also, so the frequency shifts in real experiment should have some dispersion, and – because of there are interactions of the Earth parts on distance $2R_E$, - minimal shift is ~3.7 Hz. As it was pointed above, at the calculations of this dispersion in [18] it seems wrong value for elemental momentum was used, but for the rest all equations of [18] are valid, so further we don't repeat its in this article.

Besides - since Eqs(18)-(20) above are valid only for static (non-relativistic) situation, when photons are relativistic particles principally, here we point only of course rather rough estimations of the initial beam dispersion (the Table 1):

Table 1. The fractions of initial vertical to the surface monochromatic beam after passing the way, $l$, and having dispersed frequency that differ from initial value more then on $\Delta \nu$ in Earth gravity field.

| $l$ [m] | H-maser (22 cm) $\nu_0$[Hz]= 1.4 $10^9$ | | | H2O maser $\nu_0$[Hz]= 2.2 $10^{10}$ | | | 1 eV (1250 nm) $\nu_0$[Hz]= 2.4 $10^{14}$ | | |
|---|---|---|---|---|---|---|---|---|---|
| | $\Delta\nu$>3.7 [Hz] | >160 [Hz] | $\Delta\nu$ GR [Hz] | $\Delta\nu$>3.7 [Hz] | >160 [Hz] | $\Delta\nu$ GR [Hz] | $\Delta\nu$>3.7 [Hz] | >160 [Hz] | $\Delta\nu$ GR [Hz] |
| 1 | 1 $10^{-8}$ | 5 $10^{-12}$ | 2.5 $10^{-7}$ | 1.5 $10^{-7}$ | 1 $10^{-10}$ | 3.8 $10^{-6}$ | 3.4 $10^{-3}$ | 1.8 $10^{-6}$ | 0.042 |
| 10 | 1 $10^{-7}$ | 5 $10^{-11}$ | 2.5 $10^{-6}$ | 1.5 $10^{-6}$ | 1 $10^{-9}$ | 3.8 $10^{-5}$ | 3.4 $10^{-2}$ | 1.8 $10^{-5}$ | 0.42 |
| 100 | 1 $10^{-6}$ | 5 $10^{-10}$ | 2.5 $10^{-5}$ | 1.5 $10^{-5}$ | 1 $10^{-8}$ | 3.8 $10^{-4}$ | 0.34 | 1.8 $10^{-4}$ | 4.2 |
| 1000 | 1 $10^{-5}$ | 5 $10^{-8}$ | 2.5 $10^{-4}$ | 1.5 $10^{-4}$ | 1 $10^{-7}$ | 3.8 $10^{-3}$ | 100% | 1.8 $10^{-3}$ | 42 |
| $10^4$ | 1 $10^{-4}$ | 5 $10^{-7}$ | 2.5 $10^{-3}$ | 1.5 $10^{-3}$ | 1 $10^{-6}$ | 3.8 $10^{-2}$ | | ~2% | 420 |
| $10^7$ | ~20% | ~$10^{-5}$ | ~0.9 (*) | ~0.5 | ~2 $10^{-4}$ | ~14 | | | ~1.5 $10^5$ |

(*) – there was the attempt to measure this GR shift 0.9Hz (R.F.C. Vessot et. al., Phys. Rev. Lett., V45, No 26, 1980, P 2081; but they used a signal filtration with "necessary" filter's band and 100 s averaging time).

So if some photon beam passes some distance in the Earth gravity, then it becomes be split on two fractions: (i) – the initial photons with initial frequency, and (ii) – the photons having random frequency and phase; that is in contrast to GR prediction when all photons frequencies become coherently shifted on equal value. So a technique that will allow selecting some non- coherent part in a photon beam can to detect the randomness of gravity force and to support by such a way suggested informational concept.

## 3.2. The Beginning

### 3.2.1 "In Beginning was the Word…"

Now there exists a number of physics theories that relates to the problem of Universe creation, most popular - "Big Bang hypnotises" where it is suggested that Universe was created from a point having singular dense and temperature. But - even as some hypnotises - these theories haven't any reasonable conjectures – from where (how) did this Point with starting energy $10^{85} – 10^{90}$ MeV appear?

Generally speaking, in the informational concept, yet Zero element contains all information about our Universe "far before Beginning", as a partial true [negation] statement "theirs Universe doesn't exists "now"". The logical singularity of this statement is infinite so corresponding information was enough for creation of the Universe at a "Big Logical Bang»; as well as - for further Universe's evolution. Necessary information in the statement *existed in an implicit form, but that isn't problem* – above an example was given, when an implicit information in a system of axiom can be successfully "evolved" into, may be infinite, tail of theorem, problems, calculations, etc., of some theory; including – such a process can be carried out automatically in a computer if some corresponding code starts up. So we cannot to principally reject the version "In the Beginning was the Word and the Word was…" now.

### 3.2.2. "Material scenarios"

But in the informational concept it's possible to suggest "more material", then some "paternoster version", version of the Universe creation. And here aren't too many problems also, if one has in mind again the property of the information that was many times mentioned above – *a fixed information implicitly contains* infinite information, including, in some specific cases/ contexts - a *large dynamic information*.

For example let "somebody" build some short fixed information – closed loop of 6 FLEs, let – in the state "1". Then after a small action on a FLE the flipping process in the loop starts run resulting in appearance in Space of (probably smallest in Matter) a particle having – see Eq.(1) - the mass $\sim 10^{22}$ MeV, i.e. – Planck mass particle (PMP). If our Universe "weighs" $\sim 10^{90}$ MeV then after $\sim 10^{20-30}$ s it can be created as whole. Corresponding time seems be too long, but if the process will be, e.g., exponential, say – doubling of PMP number after next step, then it is enough $\sim 200 – 300$ steps (or $\sim 10^{-40}$ s) to finish.

Another example - let fixed information – a particle having Compton length $\sim 10^{33}$ m and so mass $\sim 10^{-73}$ g exists. Then with non- zero probability an accidental brake of the loop can lead, in principle, to creation of "necessary set" of PMP; another example of fixed information – some closed loop of hexagons (i.e. - PMP structures), which, if packed in 3-dimensions, will occupy $\sim$ μm region in Space, at that – since all FLEs of individual PMPs are affiliated to neighbours PMPs – the mass of such a structure can be equal zero at all; etc.

So there exist many possible scenarios for Creation and the problem remains – "which is utmost "easy"? But the answer is maybe interesting first of all for people, who are going to make some own Worlds. For us now the choice of scenario seems not too much of principle – to understand Nature on this stage it is enough to know that Creation process doesn't require something non- understandable.

## 3.3. Dark matter

In the consideration above Planck mass particles are considered as some initial "bricks" of the Universe because of PMPs are from one hand – simplest (though more simple can exist, e.g., "triangle" structure, but "natural" is a hexagon – why?) informational structures, so it require least information to be created and in some accidental process the creation of such a structures seems utmost probable; from another – since utmost maximal t-IC – its contain maximal energy/ mass if created.

So consider more properties of the PMP.
It is well known, that for a uniform black hole the radius depends on the mass in accordance with Schwarz child formulae

$$r_g = \frac{2GM}{c^2} \qquad (21)$$

The PMP mass is equal: $m_P = (\frac{\hbar c}{G})^{1/2} = \frac{\hbar}{l_P c}$ (22)

Substituting the mass value in Eq.(21) we obtain that radius of PMP, $l_P$, is less then the Schwarzschild limit and so its should be the black holes and so its don't interact with the externals by any force besides gravity.

Though that follows evidently also from utmost simple PMP structure - here is no place to write some additional information about the particle - all PMP's t-FLEs are "significantly universal", i.e., - serve only to interact with e-FLEs to exist/ move in the Space or to interact with gravitational e-FLEs' rims caused by an external mass's t-ICs.

So PMPs interact with the "Matter particles" and other PMPs through the gravitation only. To estimate the cross section of this interaction, assume that effective interaction starts from the binding energy 1 eV. Then the distance when, e.g. a PMP, can be captured by other PMP, is equal $\approx 2 \cdot 10^{-7}$ m, so the capture cross section ($1.3 \cdot 10^{-13}$ m$^2$, or $1.3 \cdot 10^{15}$ barn) is rather large. But in reality it is rather small value, because of: (i) - most of the particles in any reference frame have the energies well above 1 eV, and (ii) - the density of PMP is extremely low, $\sim 10^{-18}$ of the baryons density, i.e., $\sim 10^{-18}$ m$^{-3}$; when (iii) - "usual" particles, e.g. - baryons – are practically transparent for PMPs (the distance for "1eV potential energy" for, e.g., proton, is equal $\sim 10^{-26}$m when proton's Compton length is $\sim 10^{-16}$m ). So it is rather possible that the PMPs don't interact with the matter, including another PMPs, practically.

But if the interaction occurs, then to little (6-7 $l_P$) length PMP-code another code should be added. It is not impossible that in this case the radius of (PMP+ particle) system exceeds the Schwarzschild limit and the (PMP + particle) system becomes non – black hole system which should decay: $PMP + particle \rightarrow N$ particles having total energy $\approx 10^{19}$ BeV.
So PMP particles can be the candidates for the "dark matter" constituents when the reactions of the PMP particles with the matter can be some source of the cosmic rays; such a process could be rather intensive at Beginning when the distances between particles were comparatively small and ~ 20%-25% of PMPs could transform into "usual" matter.

*3.4 The electricity: static solution*

The electric force is rather similar to gravity - both potentials are as 1/r, if some bodies interact then in reality when the interactions of separated particles occur, etc.; except, of course, that gravity is much weaker that electric and that electric force can act as attraction and as repulsion. So it is rather reasonable to conjecture that the equations for the potentials should be similar also, but the probability of electric interaction should be larger. So for the electric coincidence current we can obtain an analogue to Eqs. (12), (13) (for a couple of particles with the elementary charge, *e*) the equation:

$$N_{cc21} = \frac{m_1 c^2 \cdot 2\pi r W_{1E}}{4\pi r^2 \hbar} P_E \frac{m_2 c^2}{\hbar} 2\tau_E, \qquad (23)$$

where $W_E$ – is the "electric rim" width, $P_E$ – the probability of the interaction if through 2-particle a rim of 1-particle passed, $\tau_E$ – the "passing'" time. Under rather plausible conjectures that: $W_{1,2E} = \alpha^{1/2} \lambda_{1,2}$, where $\lambda$ is the Compton length of a particle; $\tau_E = W_{2E}/c$; $P_E = 1$; $\alpha$ - the fine structure constant, we obtain from Eq. (23) that electric potential energy is

$$U_E = h_D \cdot N_{cc21} = \frac{\alpha \hbar c}{r} = \frac{e^2}{4\pi\varepsilon_0 r}, \qquad (24a)$$

and for EM force in the statics obtain

$$f_{E0} = \frac{dP}{dt} = N_{cc21} p_0 = \frac{e^2}{4\pi\varepsilon_0 r^2}$$
$$= \frac{q_1 q_2}{4\pi\varepsilon_0 r^2}. \qquad (24b)$$

(the second term in Eq.(24b) is for arbitrary charges).

Note that, as what was obtained above for gravity, there should exist the "electrical time dilation" of the tied electrical structures, e.g. – for atoms. So, e.g., in the ($\mu-meson+electron$) "atom" $\mu-meson$ should live longer then in free state.

### 3.5. The forces at the movement

### 3.5.1. The electricity

Consider a straight movement of a charge $q$. As it was pointed out in [1], since Compton length is proportional to Lorentz factor, $\gamma$, the "electric coincidence" current doesn't, in some sense, depend on the speed of a particle, so "electric charge is relativistic invariant" (in fact, the relativistic invariant is square root of fine structure constant, $\alpha$) – analogously to the spin. Besides note, that in the "charge section» of a particle's code, $W_E$, the main part of t-FLEs aren't "universal" in the particle code an so aren't used in localisation of the particle in Space. Then from Eq.(24a) obtain the equation for electric potential of the charge $q$, $q=Ne$:

$$\varphi(r) = N \frac{\hbar \alpha^{1/2} c}{r} \equiv \varphi_0 \qquad (25)$$

In reality the potential is the flux of e-FLEs in a point $\vec{r}$, where the e-FLEs are grouped in "bunches" having the width $W_E$ if a charge is in the rest. If the charge moves with the speed $\vec{v}$ relating to a rest reference frame, then the potential in this frame is:

$$\varphi(\vec{r}) = \varphi_0 \frac{1}{\gamma(1-\beta\cos(\theta))} \cdot \gamma = \varphi_0 \frac{1}{(1-\beta\cos(\theta))}. \qquad (26)$$

where second factor in Eq.(26) is the Doppler factor for the e-FLE bunches, third factor - $\gamma$ - takes into account that the time for the bunch interception of the charge section of second particle in point $\vec{r}$ increases, $\theta$ is the angle between $\vec{v}$ and $\vec{r}$.

It is evident that Eq. (26) is well known equation for retarded Lienard – Wiechert scalar potential. If we remember, at a movement through Space a rim's e-FLE is, nevertheless, some informational current and must also "spend" some of it's flips on the shift in Space because of it's source, i.e.- of the charge, - movement, then quite naturally obtain that electric potential is 4- vector, and it is for a moving charge:

$$A_\mu = (\varphi, \vec{\beta}\varphi) \equiv (\varphi, \vec{A}) \qquad (27)$$

Note also, that Eq.(26) contains some evidence that rather possibly the "the charge photons" are the "rims" when "usual photons" (which occur at, e.g., an atom's transitions) are something as corpuscles; because of in the last case Lienard – Wiechert scalar potential would be proportional at least to square of Doppler factor owing to the aberration/ beaming.

When some another charge, $q_2$, moves in the same rest frame with a speed $\vec{\beta}_2$, then the coincidence time current in this charge will be equal: $j_{tc\mu} = \frac{1}{\hbar} q_2 A_\mu$ and further, the rate of the information change on the charge's 2 way is

$$\frac{dI}{dt} = j_{tc\mu} \cdot \frac{dx_\mu}{dt} = \frac{1}{\hbar}(q_2\varphi - q_2\vec{\beta}_2\vec{A}) \qquad (28)$$

From Eqs. (10), (28) we again obtain that Lagrangian of moving charged particle is the rate of the information change on it's way in Space, with negative sign.

So we can calculate the parameters of a charged particle movement by using well known equations of classical *ED, having in mind that eventually we only "decode" by these equations some informational processes in the matter and it is possible that further development of the inform concept will let to obtain more data about these processes.*

### 3.5.1.1 *But we can also obtain these results by following way*

So for the electricity "charge photons", i.e. – flipping e-FLEs, that are groped in the bunches, but are independent to some extent, these photons have (and – that is possibly true for the "gravitons" in section 3.1) the two-component structure – see Fig.6:

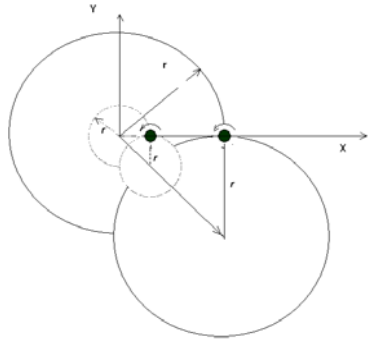

Fig. 6. Propagation of a charge photon (two positions are shown). The t- axis is directed on the reader. The large black point is the flipping FLE of the charge photon.

(1) "R- photon", i.e. – the FLE –rim that propagates in Space (in Fig.6 – in the plane (*XY*); and
(2) "F(orce)- photon", i.e., - the FLE which flips in some (in Fig.6 – along  *X*- axes) direction. F-photon, in fact, seeks for to be "usual" particle – i.e. – to have a closed-loop code in the spatial plain, but because of that it's Compton length is ever-changing – this photon turns out to be a restmassless particle. But it has a moment – just as the flipping FLE in the loop of usual material particle.

Besides his moment, $p_0 = \hbar/r$, F-photon has own angular momentum, $\vec{M}_F = I\vec{\omega}$, where *I* is the moment of inertia of the FLE, $\omega$ is angular speed of flipping FLE. In the statics the angular momentum and the angular speed are directed normally to (here) X-axes.

At the interaction with a "charge FLE" of "receiving" particle, F-photon transfers his angular moment, as well as the moment $p_0$, to this FLE. It is rather probable, that, for example, if the angular moments of F-photon and the FLE are directed identically, then a repulsion occurs with increasing of the t-IC of the particle; the situation when the angular moments have opposite directions leads to the attraction (slowing down the t-IC, mass defect, etc.)

*If "radiating" charge moves*

The case when "radiating" charge moves is shown in Fig. 7.

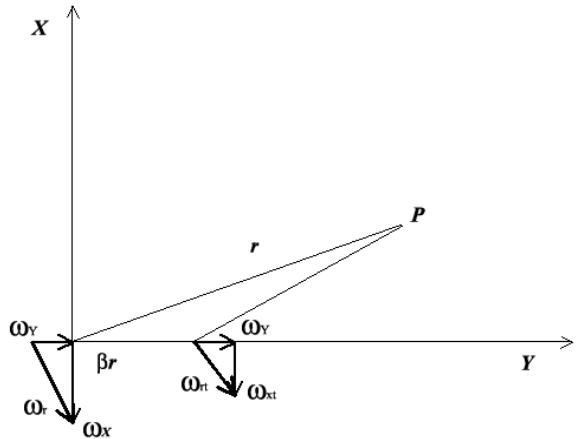

Fig. 7. The case when "radiating" charge moves. Radiating charge is in "retarded" point (0,0,0). Receiving charge is in the point *P*. Speed of "radiating" charge is equal β. Two vector sums of F-photon's angular speed are shown: left is for the static case, right – for moving charge.

The angular speed of the F-photon's FLE, $\omega_r$, we can decompose on *X* and *Y* components, as that is shown in the Fig.7. In the static as that was pointed out above, resulting vector is directed at right angle to the line between the charges. But if radiating charge moves, it is not necessary at the movement in the X direction to make all flips, so $\omega_x$ decreases on the part equal to *β,* thus resulting angular speed becomes be equal

$$\vec{\omega}_{rt} = \vec{\omega}_y + \vec{\omega}_x(1-\beta) \qquad (29)$$

and transferred momentum is equal

$$\vec{p}_{0rt} = p_{0t}(\frac{\vec{r}}{r} - \vec{\beta}) \qquad (30)$$

Here the symbol for the elementary momentum is changed, because of F-photons of static and moving charges possibly aren't identical. And here we assume that momentum of F-photon changes as the momentum of "usual" photon, i.e. it becomes be proportional to the Doppler factor. Eventually for transferred momentum we have:

$$\vec{p}_{0rt} = p_0(\frac{\vec{r}}{r} - \vec{\beta})\frac{(1-\beta^2)^{1/2}}{1-\beta Cos(\theta)} = \frac{\hbar}{r}(\frac{\vec{r}}{r} - \vec{\beta})\frac{(1-\beta^2)^{1/2}}{1-\beta Cos(\theta)} . \qquad (31)$$

Taking into account the equation for the flux of the rims, Eq.(26), and that the flux of FLE *in the bunches* should also be transformed by Doppler factor, obtain the equation for the electric field of a moving charge:

$$\vec{E}(\vec{r}) = \varphi(\vec{r})\frac{1}{r}(\frac{\vec{r}}{r} - \vec{\beta})\frac{(1-\beta^2)}{[1-\beta Cos(\theta)]^2} , \qquad (32)$$

that is well known from corresponding textbooks.

*Usual photons*

"Usual" photons, i.e. that appear at, for example, atomic transitions, obviously differ from the "charge" ones – for example, when charge photons can act by repulsion and attraction, depending on the sign of receiving charge, usual photon acts only as repulsive particle.

So for the usual photon we can suggest another model that is shown in Fig. 8.

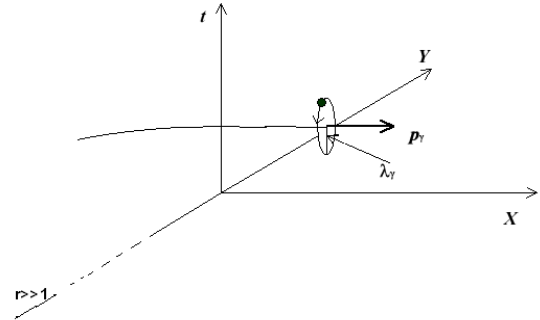

Fig. 8. Propagation of an "usual" photon. The large black point is the flipping FLE of the photon. Compare with Fig. 2.

In the model it is assumed that "usual" photon appears at some impact on a charge photon with transferring to it the momentum, $\vec{p}_\gamma >> \hbar/r$, at that the photon "forgets" it's previous history and becomes to be as something like to usual particle; but travelling along a straight line remains be "restmassless".

*3.5.2 The gravity*

Since the gravity end electricity are some resembles, the gravity should have the vector potential also. And an analogue between the forces in the statics keeps also at uniform and straight movement. Indeed, if one choose for a couple of charged masses the magnitudes $m_1, m_2, q_1, q_2$, so that the 2-bodies system is in a balance at static conditions, then at any *r* and in any inertial frame that moves relating to the rest frame with any given speed, *v*, and at any angle between $\vec{v}$ and $\vec{r}$, this balance will conserve.

But electric charge is relativistic invariant, when that is non-correct for a moving mass – it is proportional to the Lorentz factor.

To overcome this in some works ([23], [24]) the suggestions that the mass is relativistic invariant also or that there exists some "Gyron field" were made. In the informational approach it is reasonable to conjecture that when in the statics the gravity potential is proportional to the full time IC, at a movement a part of the t-IC "is spending" on the movement, and only the difference of time and space currents, $j_t - j_x = j_t(1-\beta^2)$ is used for the radiation of "gravity rims". At that the t-IC of other – "receiving"- body is capable to interact with "rims" without losses. So for gravity field the scalar potential becomes as like the electrical one, but with additional factor $(1-\beta^2)$:

$$\phi_G(\vec{r},t) = \frac{G(\gamma m_1)(1-\beta^2)}{r} \qquad (33)$$

$$A_{\mu G} = (\varphi_G, \vec{\beta}\varphi_G) \equiv (\varphi_G, \vec{A}_G) \qquad (34)$$

In this case the two charge/masses straight movement problem becomes solved [1], v5; including – it turns out to be that gravitomagnetic interaction results in the parallel mass currents repulsion – when parallel electric currents attract.

But it should be noted, that the model above is a "Newtonian version", when it is experimentally strictly proved that that Newton law isn't correct in relativistic case and true version must take into account General Relativity approach. For this first of all seems be necessary to understand – what does a FLE make when being accelerated?

# 4. DISCUSSIONS AND CONCLUSION

Of course above only some initial physical model is presented. First of all – as it was pointed out in [1] – it seems that any force mediator should be a composition of at least two fermions, which "live in contrast times", i.e. a composition "fermion + antifermion" with parallel spins, including, e.g., the photon. For strong force that's well known - practically all mesons are combinations "quark + antiquark" when gluons are even more complicated. For the photon there exist some hint already in classical electrodynamics - such a components are electric and magnetic fields, when the equations for the curls of electric and magnetic fields have opposite signs of temporal derivatives.

Besides note, that the conjectures in sections 3.1, 3.4 can be developed quite natural way – it seems that a particle "radiate" rims having not only gravity and electric loop code sections, but the code as a whole. So in fact the copies of the particle are spreading in the space-time. Since FLEs in these copies are shifted on the flip time, the copies aren't closed and so they aren't cycled time ICs and correspondingly they are massless (virtual) "particles". If such a "particle" is impacted by some way and the impact's momentum is large enough to close the code, then a real particle appears. But owing to the FLE's logical structure, it seems that such a process can happen only if a pair "particle + antiparticle" appears and we should again to assume that this structure is rather complicated, when a gravity force mediator perhaps can be at least two–component being also and the "single be-stable FLE" gravity model in section 3 above should be possibly modified.

The suggested model's apparent shortcoming is that from the consideration above follows that at a movement a particle's parameters *change really*. E.g., an increase in speed (e.g. – from zero value) under acting upon the particle by some force in some reference frame leads to the decrease of the particle's size (the particle's loop code's length) and so – to the increase of the t-IC and so - of particle's mass, etc., *in this reference frame*. But one can act also on the observer to boost him up to the same speed and in new reference frame the particle should have the parameters as in the first frame, but without any action on the particle.

So the frames appear as aren't equivalent, what formally violates the relativity principle. But here we can note, that once some reference frame is established then all physical processes in any other frame will be correctly described – including, e.g. considered test problem - the forces between moving particles will be determined /calculated as such the forces in corresponding "rest frame". And if we think, e.g., that our Universe is some finite structure and appeared as a result of some Bang, then we should admit that such a frame exists in the Universe – that is the frame where the sum of momentums of all particles in Universe is equal zero and the sum of energies (total Universe's t-IC) is minimal.

Besides, there is some reason to support presented here model – it seems that it isn't impossible that the relativity principle is "virtual", i.e. is only a sequence of reversibility in time of informational codes in the Matter. And now at least the twin paradox becomes be resolved – in any case the twin-traveller will be younger then the twin-homebody - to save some years is necessary to spend some energy.

Nevertheless, even initial suggested model allows to make more understandable too many problems in physics and thus seems as worthwhile for further development.

Besides what is surely clear now already - that the Information is utmost fundamental and "utmost infinite" entity, which exists as a Set. In our World (Universe), as well as in possible other Worlds, under sequence of informational exchanges between Matter's (and possibly the rest of the Set's) elements at each step (at an ISs interaction) next "instant" World picture is created from ready bill of fare of "always known" possible variants of the evolution; these instant random pictures we call the "Matter", "the objective reality".

However note that there exist two notions of the randomness/ probability concepts - its can be "objective" and "subjective"; what corresponds, on another words, to two cases "nobody knows [the result]" and "I don't know". For instance: when a coin is in the air, nobody knows on which side this coin will fall on the ground and knows only that the probabilities of heads or tails are ½. But when the coin has fallen on the ground – somebody knows upper coin's size, when another, who don't see the coin, know only "subjective" probabilities, which remain be equal ½.[*]

When an electron passes through a slit in a screen, the point where it hit on a detector seems as having some random position. But we already know that all about anything in our Universe was exactly known "far before Beginning". In this case "far before" there was the exact true information: "there doesn't exist an Universe where after $\sim 1.4 \cdot 10^{10}$ years after Beginning in given room with given experimenters given (say - hundredth) electron will escape from given cathode and will hit in given detector's point" – so in principle there is no "objective" information – only "subjective" exists. On another hand for human's perception this information appears as "objective", because of the humans till now can not read such information in the Set, and so we have here "nobody knows" case, i.e. observe "objectively" random electrons' ways. But the Universe computer is simple; so we can obtain some "objective" information about the electron movement when study, e.g. in this case, diffraction patterns after passing of an ensemble of electrons through a slit.

---

[*] There can exist, of course, some other cases of "purely subjective" probabilities also. E.g.- if somebody (as, e.g., the authors) has a good health, he can think that at least in next couple of years all will be well with large "objective" probability. But because of some external peoples, it can be possible that this probability can become essentially less on another "subjective" value.

Now a rather popular line in physics development appeared – the development of "the Theory of Everything", when this theory should unite four "fundamental" forces that are considered in physics and are studied in corresponding physics branches. The attempts to create ToE became more intensive after successful development of the theory that "united" EM and weak forces.

But it is evident, that such a ToE can not be the theory of everything, e.g., - at new experiments with great probability new data beyond the today Physic's laws system will be discovered; what will require introducing into Physics corresponding next forces and further - development of the "Theory of next Everything", and so on. Such a process can be rather long, but in reality, as it follows from this information concept, eventually the ToE will be the theory of the Set "Information"; and, notwithstanding the "wildness" of this Set, It is, nevertheless, some mathematical object which can be studied already by using existent instruments – the set theory, the theory of the language, the synergetics, the cybernetics, etc.; when the Physics should be used to support these sciences by *new experimental data* and to "decode" their - sometimes rather abstract – results into the form being more convenient for cognition by human consciousness.

Besides note, that true ToE rather possibly should study also Set's subsets which aren't considered in the Physics now, including the interactions of "material" and "non-material" (though perhaps exchanging only by true information) things. For example – as a first step it would be interesting to explore the problem of human consciousness stability. Since consciousnesses operate sometimes with false information, they should be some elements of some another - non-material - subset. For a consciousness to be a stable structure in the Set there are at least two possible variants: (i) – the "consciousness (C-) subset" is constituted from corresponding stable "consciousness FLEs" and so C-subset is some self-dependent structure, or (ii) – a stable consciousness can exist only on a stable (for humans – on material brain) matrix. What is true?

But in this case the Physics will become the Metaphysics again…

*Acknowledgements*: The author would like to thank the members of a number of Inet physical forums for helpful discussions[1].

---

[1] In particular we thank the members of Moscow State University Physical Department forum - http://www.dubinushka.ru , forum; the thread "The information and the Matter" (in Russian), April 2006 – May 2009)